\documentclass[preprint,aps]{revtex4-1}
\usepackage{color}
\usepackage{mathtools}
\usepackage{array}
\usepackage{tabularx}
\usepackage{diagbox}
\usepackage{tikz}
\usetikzlibrary{decorations.pathmorphing}
\usepackage{bm}
\usepackage{amsmath,bm,amssymb,amsfonts,dcolumn,color,graphicx,latexsym,epsfig}
\usepackage{rsfso}
\usepackage{hyperref}
\hypersetup{
colorlinks=true,
linkcolor=magenta,
filecolor=magenta,      
urlcolor=blue,
citecolor=blue,
}
\usepackage{multirow}
\usepackage{natbib}
\usepackage{float}
\usepackage{booktabs}
\usepackage{pifont}
\usepackage{mathrsfs}
\usepackage{caption}
\usepackage{caption, threeparttable}
\begin{document}

\title{Novel regular black holes: geometry, source and shadow}

\author{Anjan Kar and Sayan Kar}
\email{anjankar.phys@gmail.com, sayan@phy.iitkgp.ac.in}
\affiliation{Department of Physics, Indian Institute of Technology, Kharagpur 721 302, India}

\begin{abstract}
\noindent We propose a two-parameter, static and spherically symmetric regular geometry, which, for specific parameter values represents a regular black hole. The matter required to support such spacetimes within the framework of General Relativity (GR), is found to violate the energy conditions, though not in the entire domain of the radial coordinate. A particular choice of the parameters reduces the regular black hole to a singular, mutated Reissner–Nordström geometry. It also turns out that our regular black hole is geodesically complete.  Fortunately, despite energy condition violation, we are able to construct a viable source, within the framework of GR coupled to matter, for our regular geometry. The source term involves a nonlinear magnetic monopole in a chosen version of nonlinear electrodynamics. We also suggest an alternative approach towards
constructing a source, using the effective Einstein equations which arise in the context of braneworld gravity. Finally, we obtain the circular shadow profile of our regular black hole and provide a preliminary estimate of the metric parameters using recent observational results from the EHT collaboration.   
\end{abstract}


\maketitle

\newpage

\section{Introduction}
 \noindent In recent years, the detection of gravitational waves from binary black hole mergers, black hole--neutron star binaries as well as neutron star-neutron star binaries \cite{Ligo1, Ligo2, Ligo3, Ligo4} have boosted research in gravitational physics. On another front, imaging observations on shadows due to the strong gravitational lensing around black holes in different galaxies, have provided useful information about such supermassive compact objects \cite{Akiyama1, Akiyama2, Akiyama3, Akiyama4, Akiyama5, Akiyama6, Akiyama7, Akiyama8, Akiyama9, Akiyama10, Akiyama11, Akiyama12}. Since Einstein's theory of General Relativity (GR) is, by and large, the most acceptable theory of gravity today, the known rotating vacuum solution of Einstein's equation, i.e. the Kerr solution, along with its properties, must explain phenomena related to such compact objects having horizons. It is, by now known, that imaging observations have more or less confirmed that the compact objects have a Kerr-like behaviour within a small uncertainty. However, to describe such compact objects, other solutions with horizons, predicted by different theories of gravity cannot be completely ignored. This implies a sort of `degeneracy' in the sense that `many models' can explain available data successfully. Breaking the `degeneracy' will therefore require more observations as well as newer theoretical models.  
 
 \noindent On the other hand, from a theoretical perspective, the Hawking-Penrose singularity theorems \cite{Hawking} with certain assumptions (energy conditions among them) have indeed proved that black hole solutions in GR possess a spacetime singularity(geodesic incompleteness). This is normally projected as a sign of inconsistency of classical GR and one would like to get rid of or avoid such singularities in some way, classical or quantum. However, one does not quite have a universally accepted resolution of this `singularity problem'.  

 \noindent Thus, from an observational perspective (as mentioned above) as well as the desire for well-behaved solutions in a theory of gravity, one is motivated to search for non-singular (regular) solutions in GR as well as in modified gravity theories.    

\noindent Regular black holes are known potential candidates for such non-singular solutions.  They may violate the Strong Energy Condition (SEC) and, therefore, can circumvent the singularity theorem \cite{Ansoldi, Zaslavskii}. As classical GR fails at the singularity, Sakharov \cite{Sakharov} and Gliner \cite{Gliner}, invoking quantum aspects, indicated that a de Sitter core could replace the singularity inside the horizon. Based on this idea, in 1968, Bardeen first proposed a static spherically symmetric regular black hole spacetime \cite{Bardeen}. Since then, numerous proposals for regular black holes  \cite{Bardeen2, Bardeen3, Hayward, Roman, Dymnikova, Dymnikova1, Ayon1, Ayon3, Frolov, Frolov2, Frolov3, Balart, Bronnikov, simpson, Carballo2, Carballo3, Bambi, Ghosh, Sajadi, Roy, tapo1} have come up. The first proposal for the matter source of a regular black hole of the Bardeen type was made using a nonlinear electric or magnetic monopole \cite{Ayon1, Ayon2, Ayon3, Ayon4, Ayon5} i.e., a nonlinear electromagnetic field Lagrangian $(\mathcal{L}(\mathbf{F}))$ coupled with gravity was found to be capable of generating the matter required to support such a spacetime. Subsequently, various regular black hole models have emerged wherein a nonlinear electromagnetic field Lagrangian is employed as a matter source \cite{Balart, Fan, Zhong, Bronnikov3, Bronnikov4, somalic}.

 \noindent Though popular of late, there are several criticisms of a regular black hole. First, we still do not understand properly the nature and dynamics of the source Lagrangian $(\mathcal{L}(\mathbf{F}))$. In other words, we have much to learn about the so-called nonlinear Maxwell-like equations. Second, each of the constructed regular black hole models have a different source Lagrangian $(\mathcal{L}(\mathbf{F}))$, which means we do not have a single, well-defined Lagrangian which can take on different functional forms when solutions for the input fields $(\mathbf{F})$ are used to describe the required matter for regular black holes. In addition, the mass inflation instability \cite{Poisson1, Poisson2} also appears to be a serious issue.  The mass function increases exponentially under a perturbation at the inner horizon. Recently, significant efforts have been put forward to resolve the problem \cite{Bonanno, Carballo}, but the issue remains open.

 \noindent Despite the shortcomings related to any research on regular black holes, we prefer to be hopeful and look for directions and approaches that might potentially provide novel insights about their properties.  Let us now state our proposal to build novel spacetimes representing regular black holes. Our work is motivated by the Einstein-Rosen bridge construction  \cite{Rosen}, where the line element may be specified as,
\begin{equation}\label{eq1}
    ds^{2}=-\Big(1-\frac{b_{0}^{2}}{r^{2}}\Big) dt^{2}+ \frac{dr^{2}}{1-\frac{b_{0}^{2}}{r^{2}}}+r^{2}(d\theta^{2}+\sin^{2}{\theta}d\phi^{2})
\end{equation}
Here, $b_0$ is a metric parameter with dimensions of length, $r=b_0$ is the event horizon and the spacetime singularity
is at $r=0$.
We choose the above singular line element and regularise the metric functions such that all curvature components and scalars become regular everywhere. Though our
proposed regular black hole does violate the classical energy conditions in some domain of the radial coordinate, it is different in its geometric structure from other known regular black holes, all of which reduce to the Schwarzschild black hole for a zero value of the regularizing parameter. In a sense, we are able to show that it is not always necessary to `regularise around Schwarzschild'. Further, the process of regularisation about a solution built out of energy-condition violating matter may lead to a situation where regularisation not only remedies the singular character but also improves the status of the energy conditions {\em vis-a-vis} their satisfaction/violation. The ensuing sections of this article provide examples addressing these points.

 \noindent Our article is organised as follows. In Section \ref{second}, we propose our line element and provide a detailed study of its geometry, the energy conditions and check its non-singular character. In Section \ref{III}, we construct possible sources for the geometry using  (i) nonlinear electrodynamics
 and (ii) braneworld gravity. Section \ref{IV} discusses the circular shadow of our regular black hole and checks its viability with reference to observational data. Finally, in Section \ref{V}, we conclude with some future directions.

\section{The proposed geometry and its features}\label{second}

\noindent As stated in the Introduction, we propose a non-singular, two-parameter, static and spherically symmetric spacetime which may represent a regular black hole. Developing on the idea briefly outlined above, we  write down a line element of the following form:
\begin{eqnarray} \label{eq2} &&ds^{2} = -\Big(1-\frac{b_{0}^{2}r^{2}}{(r^{2}+g^{2})^{2}}\Big)dt^{2} + \Big(1-\frac{b_{0}^{2}r^{2}}{(r^{2}+g^{2})^{2}}\Big)^{-1}dr^{2}  +r^{2}\Big(d\theta^{2}+\sin^{2}\theta d\phi^{2}\Big).
\end{eqnarray}
 Here, $g^{2}$ is the regularizing parameter and the choice of the metric functions is largely inspired by earlier constructions due to Bardeen \cite{Bardeen}. Note that our metric is not similar or related to the well-known regular black hole spacetimes of Bardeen, Hayward, or Simpson-Visser \cite{Bardeen, Hayward, simpson}, all of which reduce to the standard Schwarzschild line element when the regularizing parameter vanishes. Instead, corresponding to the parameter value $g^{2}=0$, our metric becomes a mutated, singular Reissner–Nordström (RN) solution with an `imaginary charge' and a vanishing mass parameter. Further, the spacetime metric is asymptotically Minkowski, i.e.
\begin{equation*}
    g_{tt}\to\ -1 \hspace{0.7cm} and \hspace{0.7 cm} g_{rr}\to\ 1 \hspace{0.7 cm} as \hspace{0.7 cm} r\to\ \infty
\end{equation*}
For small values of $r$, the metric behaves like de-Sitter space, i.e.
\begin{equation*}
    g_{tt}\to\ -(1-c_{1}^{2}r^{2}) \hspace{0.7cm} and \hspace{0.7 cm} g^{rr}\to\ 1-c_{1}^{2}r^{2} \hspace{0.7cm} as \hspace{0.7 cm} r\to\ 0
\end{equation*}
  Before proceeding further, let us mention the natural domain of the coordinates:
$$r\in(0,\infty);\hspace{0.9 cm} t\in(-\infty,\infty);\hspace{0.9 cm}\theta\in[0,\pi];\hspace{0.9cm}\phi\in(0,2\pi]$$ 
One can write the above metric in terms of the following reparameterization  $g^{2}=m^{2}b_{0}^{2}$, where $m^{2}$ is a dimensionless quantity. Such a parameterization helps in calculations. The line element, with this choice, becomes:
\begin{equation}\label{eq3} 
ds^{2}=-\Big(1-\frac{b_{0}^{2}r^{2}}{(r^{2}+m^{2}b_{0}^{2})^{2}}\Big) dt^{2}+ \Big(1-\frac{b_{0}^{2}r^{2}}{(r^{2}+m^{2}b_{0}^{2})^{2}}\Big)^{-1} dr^{2}+ r^{2} \Big(d\theta^{2}+\sin^{2}\theta d\phi^{2}\Big)
 \end{equation}
The roots of the equation $g_{tt}=0$ are null hypersurfaces and represent the horizons of our black hole. Below, we discuss the nature of the spacetime and location of the horizons based on the various ranges of parameter values of $m^{2}$.
\begin{figure}[h]
\centering
\includegraphics[width=0.6\textwidth]{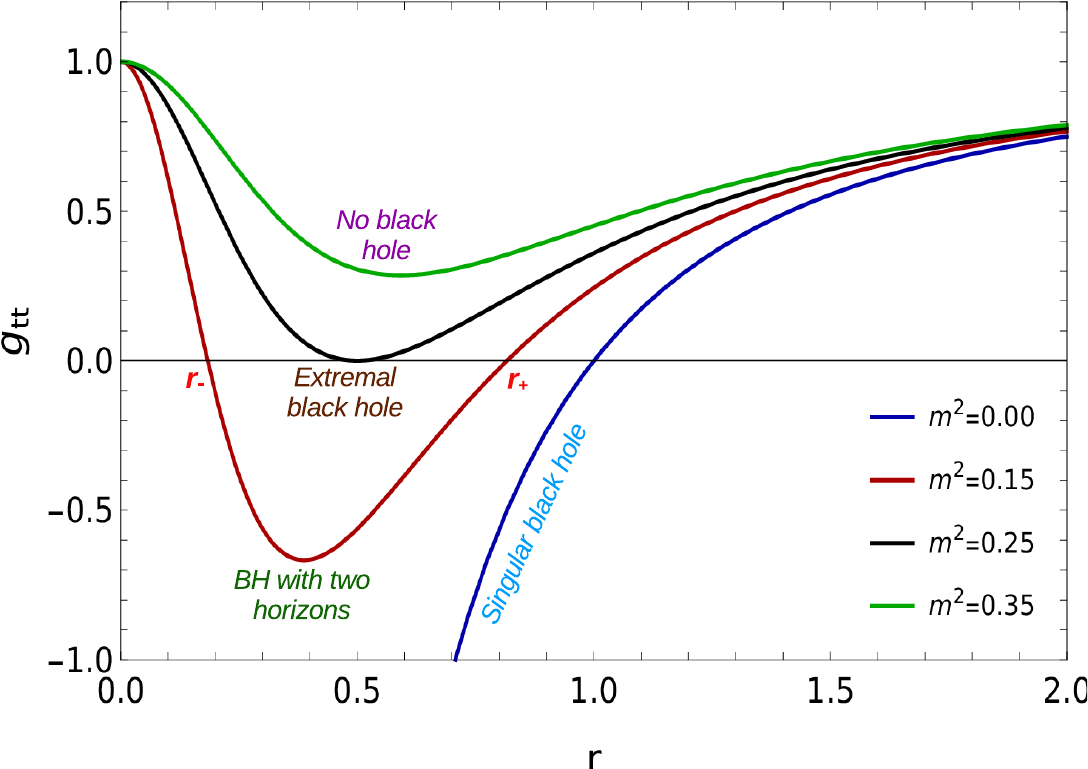}
\caption{Plot of the redshift function with $r$ for different values of $m^{2}$ }
\label{fig:horizonfunction}
\end{figure}

\begin{itemize}
    \item If $m^{2}=0$, the geometry is a singular, $M=0$ RN type solution with $-g_{tt}=g^{rr}=1-\frac{b_{0}^{2}}{r^{2}}$ having a horizon at $r=b_{0}$. It is not a regular solution. Note that the term involving the `charge'
    has a sign opposite to that for standard RN spacetime. 
    \item If $0<m^{2}\leq 0.25$, the geometry represents a family of regular black holes.  $g_{tt} =0$ has four roots, two positive and two negative. As $r\in(0,\infty)$, we consider only positive roots as horizon locations. Hence we have a black hole spacetime with two horizons. In this specified range of $m^{2}$, inner horizon$(r_{-})$ and outer horizon$(r_{+})$ vary between $0< r_{-}\leq 0.5 b_{0}$ and $b_{0}>r_{+}\geq 0.5 b_{0}$ respectively. At $m^{2}=0.25$, inner and outer horizons coalesce into a single horizon.
    This is the analog of the extremal limit, for which $b_0^2 = 4 g^2$.
    Figure \ref{fig:horizonfunction} depicts how change in $m^2$ leads to variation in the number of horizons in our spacetime.
    \item If $m^{2}> 0.25$, there are no real roots of the equation $g_{tt}=0$, i.e. we have a regular spacetime without a singularity or a horizon. 
\end{itemize}
Hence we may conclude that we have a family of regular black holes with two horizons when $0<m^{2} < 0.25$. The solution has an extremal limit similar to RN.
One may try to understand this extremal limit a little better. The original line element, in the extremal limit takes the form:
\begin{eqnarray} \label{eq4} &&ds^{2} = -\Big(1-\frac{4 g^{2}r^{2}}{(r^{2}+g^{2})^{2}}\Big)dt^{2} + \Big(1-\frac{4 g^2 r^{2}}{(r^{2}+g^{2})^{2}}\Big)^{-1}dr^{2}  +r^{2}\Big(d\theta^{2}+\sin^{2}\theta d\phi^{2}\Big).
\end{eqnarray}
The geometry is very similar to that of the extremal RN black hole. If we embed a two-dimensional $t=const$, $\theta=\frac{\pi}{2}$ slice in Euclidean space (cylindrical coordinates), the profile function $z(r)$ takes the form:
\begin{eqnarray}\label{eq5}
    z(r) = \pm g \ln \left ( \left( \frac{r}{g}\right )^2 -1 \right )
\end{eqnarray}
\begin{figure}[h]
\centering
\includegraphics[width=0.6\textwidth]{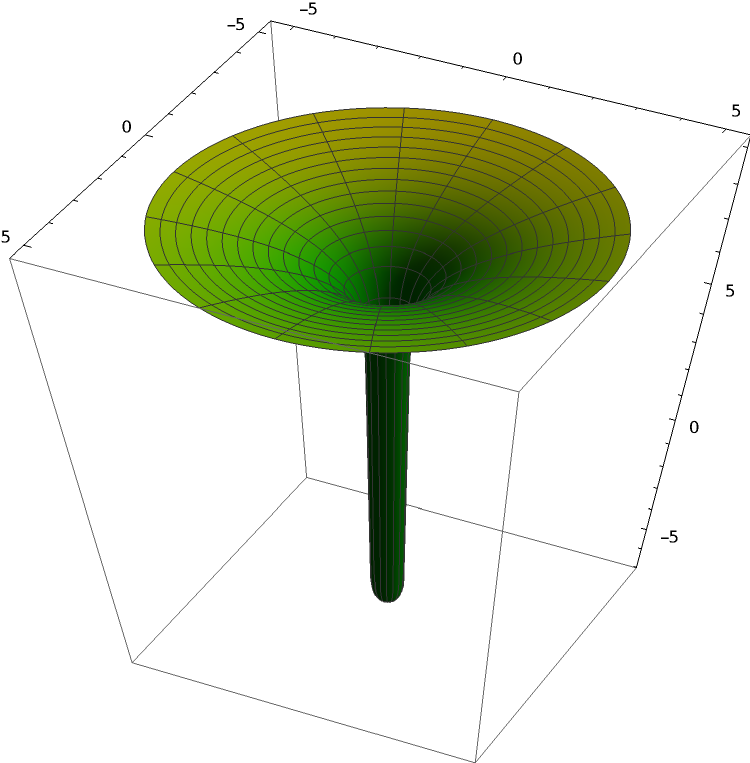}
\caption{ Embedding diagram of the extremal rbh metric shows a horn-like profile}
\label{fig:extremalrbh}
\end{figure}
which diverges to negative (positive) infinity at $r=g$ for the plus (minus) signs, respectively. The profile is like a horn or a trumpet and is similar in structure to that for the extremal RN spacetime. 

\noindent Next, we study the regularity of curvature tensors and scalars as proof of the regularity of the spacetime. We also examine geodesic completeness in our geometry and 
check the energy conditions (assuming General Relativity as the theory of gravity).

\subsection{Curvature tensors and invariants}

\noindent To investigate the singular/regular nature of our geometry, one may look for the finiteness of the Riemann and Ricci tensor components in the entire domain of coordinates. Tensor components will, of course, be frame dependent. However, individual components and their finiteness usually indicate regularity. The non-zero Riemann curvature components in the frame basis are:
\begin{equation}\label{eq6}
\begin{aligned}
     R^{01}{}_{10}&=-\frac{(g^{4}-8g^{2}r^{2}+3r^{4})b_{0}^{2}}{(r^{2}+g^{2})^{4}}\\
      R^{02}{}_{20}&= R^{03}{}_{30}=R^{21}{}_{12}= R^{31}{}_{13}=\frac{(r^{2}-g^{2})b_{0}^{2}}{(r^{2}+g^{2})^{3}}\\
     R^{32}{}_{23}&=-\frac{b_{0}^{2}}{(r^{2}+g^{2})^{2}}
\end{aligned}    
\end{equation}
Assuming $g^{2}\neq 0$, the Riemann tensor components have a finite limit: $-\frac{b_{0}^{2}}{g^{4}}$ at $r\to\ 0$. Again, for $r\to\ \infty$, all components tend to $0$, ensuring the asymptotically flat nature of the spacetime. \\  The non-zero Ricci tensor components in the frame basis are:
\begin{equation}\label{eq7}
    \begin{aligned}
       R_{00}=&-R_{11}=-\frac{(3g^{4}-8g^{2}r^{2}+r^{4})b_{0}^{2}}{(r^{2}+g^{2})^{4}}\\
       R_{22}=&R_{33}=\frac{(3g^{2}-r^{2})b_{0}^{2}}{(r^{2}+g^{2})^{3}}
    \end{aligned}
\end{equation}
 The components of the Ricci tensor also reach a finite value at $r\to\ 0$ and vanish as $r\to\ \infty$.  We can, therefore, partially conclude that there is no singularity in the metric, and it represents a family of regular black holes.

\noindent  The more important quantities to analyse are the scalar curvature invariants \cite{Dymnikova, Ayon1}. In 4D spacetime, one can construct seventeen independent curvature invariants from the Riemann curvature tensor components \cite{zi}. However, given the symmetries of our spacetime, finiteness of the three invariants -- namely, the Ricci scalar, the Ricci contraction and the Kretschmann scalar -- confirms the finiteness of all other invariants \cite{Hu}. Therefore, we only examine these three invariants.\\
The Ricci scalar is given as:
\begin{equation}\label{eq8}
    R=g^{\mu \nu}R_{\mu \nu}=\frac{12 b_{0}^{2}(g^{4}-g^{2}r^{2})}{(r^{2}+g^{2})^{4}}
\end{equation}
The Ricci contraction turns out to be,
\begin{equation}\label{eq9}
    R_{\mu \nu}R^{\mu \nu}=\frac{4b_{0}^{4}(9g^{8}-18g^{6}r^{2}+34g^{4}r^{4}-10g^{2}r^{6}+r^{8})}{(r^{2}+g^{2})^{8}}
\end{equation}
And the Kretschmann scalar is:
\begin{equation}\label{eq10}
    K=R_{\mu \nu \lambda \delta}R^{\mu \nu \lambda \delta}=\frac{8b_{0}^{4}(3g^{8}-6g^{6}r^{2}+34g^{4}r^{4}-22g^{2}r^{6}+7r^{8})}{(r^{2}+g^{2})^{8}}
\end{equation}
As $r\to\ 0$, all the curvature scalars reach a finite value and tend to zero at $r\to\ \infty$, which confirms the non-singular nature of the above geometry. 

\subsection{Geodesic completeness}
\begin{figure}[h]
\centering
\includegraphics[width=0.492\textwidth]{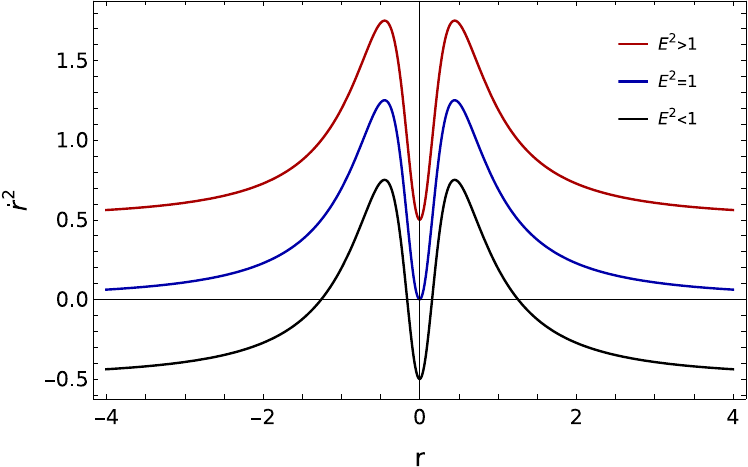}
\includegraphics[width=0.492\textwidth]{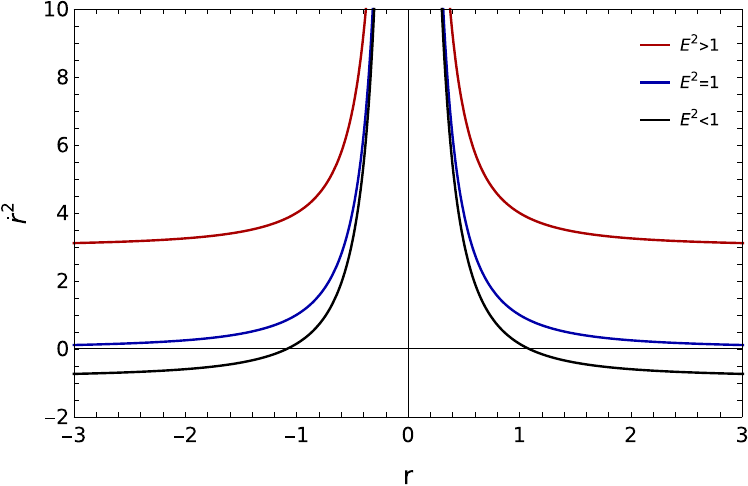}
\caption{(Left) plot of the $\dot{r}^{2}$ with $r$ of the regular metric ($m^2=0.20$) for different $E^{2}$. (Right) variation of $\dot{r}^{2}$ of the singular metric ($m^{2}=0$) shows a divergence at $r=0$.}
\label{fig:kinetic}
\end{figure}

\noindent Regularity of curvature tensors and curvature invariants are required but inadequate for testing the singular nature of a spacetime. According to \cite{Wald, Hawking, Modesto, Carballo4, Carballo5, tapo2}, the completeness of all causal geodesics is a necessary prerequisite for a regular spacetime. In this section, we look at the null and timelike geodesics in our geometry to analyze geodesic completeness. Let us start with radial timelike geodesics, which satisfy the following equation,
\begin{equation}\label{eq11}
    g_{tt}\dot{t}^{2}+g_{rr}\dot{r}^{2}=-1
\end{equation}
 where the dot represents the derivative with respect to affine parameter$(\lambda)$. Since the line element is static, we have a timelike Killing vector with the corresponding conserved quantity as $E=-g_{tt}\dot{t}$. Using this, the equation for radial timelike geodesics becomes,
\begin{equation}\label{eq12}
    \dot{r}^{2}=E^{2}-1+\frac{b_{0}^{2}r^{2}}{(r^{2}+g^{2})^{2}}
\end{equation}
The affine parameter can be defined in the following way,
\begin{equation}\label{eq13}
    \lambda(r)=\int \frac{dr}{\sqrt{\dot{r}^{2}}}
\end{equation}
This may be integrated using eq.(\ref{eq12}) to determine how $\lambda$ varies as a function of $r$. For small values of $r$, our metric acts like de-Sitter space, and if $r=0$ is a regular point, the affine parameter is finite from any arbitrary point to $r=0$. As a result, the coordinate system may be hypothetically extended beyond $r=0$, i.e. towards negative $r$ values. Again, such an extension is possible because a coordinate system is not a physical quantity. There are two alternatives for negative values of $r$: (i) one can locate a pole of $\dot{r}^2$, say at $r=-l$, where $\dot{r}^2$ diverges and if the affine parameter to reach the point is finite, then one cannot extend the co-ordinate beyond $r=-l$ (similar to $r=0$ in the Schwarzschild case), (ii) $\dot{r}^{2}$ is positive or negative and continuous up to $r\to\ -\infty$ which means for appropriate values of $E^{2}$, the trajectories of massive particles can be extended to negative values of $r$ up to $-\infty$. Thus, any divergence of $\dot{r}^{2}$ would indicate geodesic breakdown. Hence, to check geodesic completeness, our job is to check if there are any singularities in negative values of $r$ or it is regular up to $r\to\ -\infty$ with an infinite value of the affine parameter.\\ 
To perform this exercise, we have followed the prescription mentioned in \cite{Modesto}. First, one can define the effective radial potential $(V_{eff}=E^2-\dot{r}^{2})$ in terms of the parametrization mentioned earlier. We have
\begin{equation}\label{eq14}
    V_{eff}(r)=1-\frac{b_{0}^{2}r^{2}}{(r^{2}+m^{2}b_{0}^{2})^{2}}
\end{equation}
\begin{figure}[h]
\centering
\includegraphics[width=0.6\textwidth]{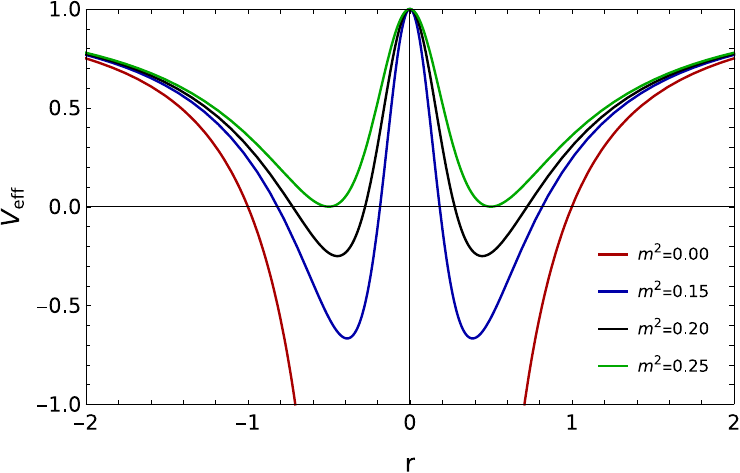}
\caption{Plot of effective potential as function of $r$ for different values of $m^{2}$}
\label{fig:potential}
\end{figure}
In Figure \ref{fig:potential}, we plot the effective potential for the extended domain of $r$, i.e. $(-\infty,\infty)$. We have a continuous effective potential for $0<m^2\leq 0.25$. A massive particle with enough energy can travel up to $r\to-\infty$. On the other hand, the singular solution ($m^2=0$), exhibits a discontinuity in the effective potential at $r=0$. As a result, in this case, the massive particle will approach $r=0$ within a finite affine parameter, indicating the presence of a singularity.
Figure \ref{fig:kinetic} shows the behaviour of the kinetic energy of massive particles. We notice a continuity in the behaviour of kinetic energy for $0<m^2\leq0.25$, (Figure \ref{fig:kinetic} (left)). However, the kinetic energy is negative if $E^2<1$. When $E^2=1$, kinetic energy reaches zero at $r=0$, and massive particles need a large affine parameter to get there. However, if $E^2>1$ a massive particle will reach $r=0$ in a finite affine parameter before moving away to $r\to-\infty$. Figure \ref{fig:kinetic} (right) illustrates the kinetic energy behaviour of massive particles in the singular geometry ($m^2=0$). It demonstrates that the kinetic energy diverges at $r=0$, regardless of the value of the total energy. As a result, within a finite affine parameter, a massive particle will reach $r=0$, where the geodesic will terminate. 
Next, the same procedure is followed for null geodesics. The right-hand side of eq.(\ref{eq11}) is now zero. Here too, there is no pathology in the effective potential or kinetic term, and null geodesics can be extended to $r\to -\infty$. However, a termination of geodesics is evident in the singular metric $(m^2=0)$ for massless particles too. 

\begin{figure}[h]
\centering
\includegraphics[width=0.88\textwidth]{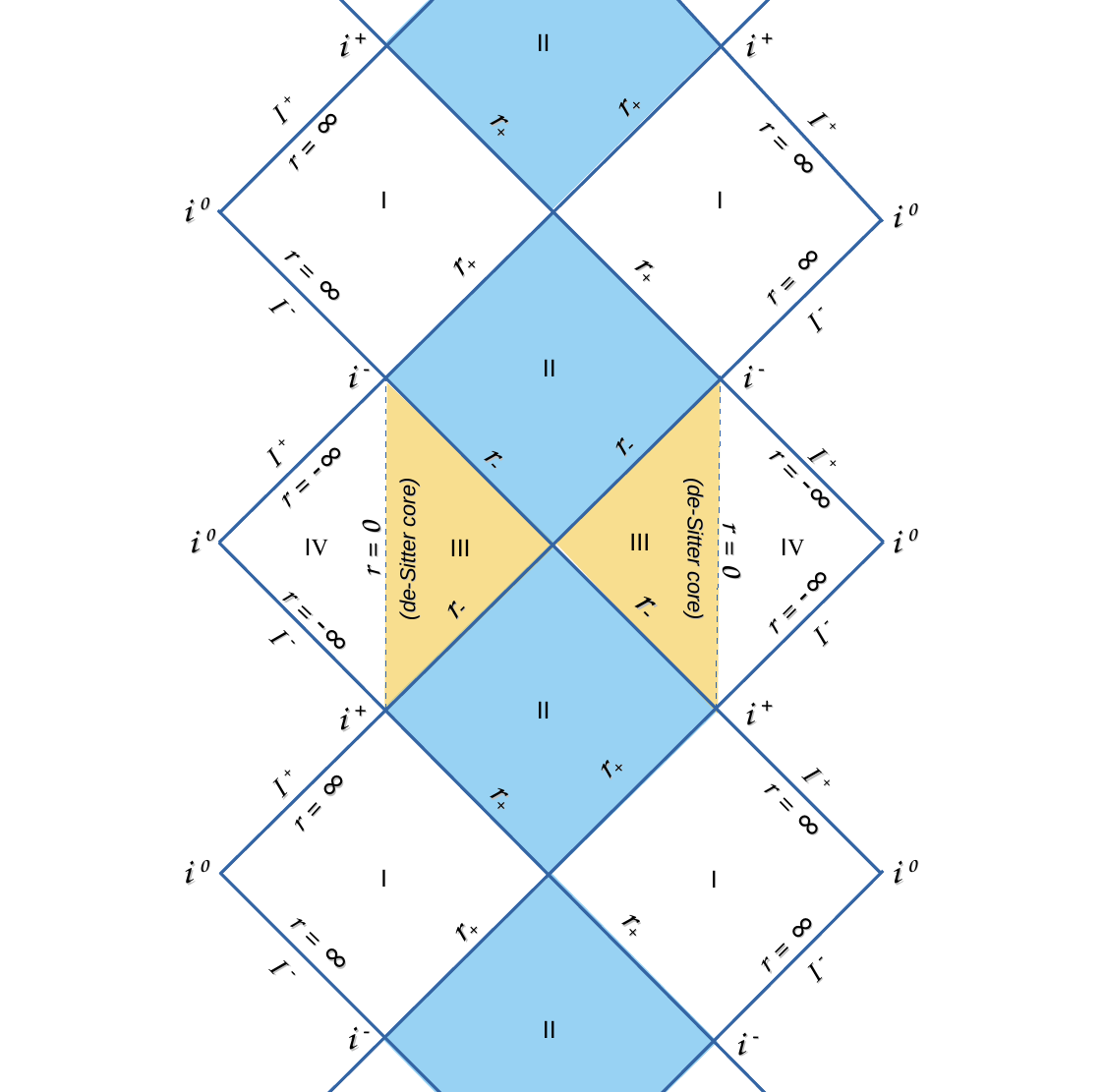}
\caption{Plot of maximal extension of our regular metric $(0<m^2\leq0.25)$ showing its geodesic completeness.}  
\label{fig:perose}
\end{figure}

\noindent To represent the causal structure of our regular spacetime, we extend the coordinates maximally \cite{Ramon}. In Figure \ref{fig:perose}. $r_{+}$, $r_{-}$ are the outer and inner horizons, respectively. The interval from the outer horizon ($r_+$) to asymptotic infinity ($r\to\infty$) is encoded in the region I. Region II lies between the outer and inner horizons. Unlike the singular RN type metric $(m^2=0)$, the core of the regular spacetime $(0<m^2\leq 0.25)$ is a de-Sitter space. Region III, therefore, depicts the range from the inner horizon ($r_-$) to $r=0$. Region IV represents an asymptotically flat space through the regular core at $r=0$ to $r\to -\infty$. In conclusion, our regular black hole geometry is geodesically complete for both massive and massless particles. However, the singular RN-type metric is geodesically incomplete. 

\subsection{Energy momentum tensors and energy conditions}

\noindent Regular black holes can avoid the Penrose singularity theorem by violating the strong energy condition \cite{Ansoldi, Zaslavskii}. Therefore, analyses of the energy conditions are important while studying any regular spacetime \cite{Bambi2, Lan}. In this section, we have looked at different energy conditions, namely the null energy condition (NEC), the weak energy condition (WEC), the dominant energy condition (DEC) and the strong energy condition (SEC). We write the diagonal elements of the required stress-energy tensor, in the frame basis, using the Einstein field equations $G_{ab}=\kappa T_{ab}$. This gives us the following
expressions for the components of $T_{ab}$:  
\begin{eqnarray}
     \rho =-\tau=\frac{(3g^{2}-r^{2})b_{0}^{2}}{\kappa(r^{2}+g^{2})^{3}} = -\frac{b_0^2 r^2}{\kappa(r^2+g^2)^3} + 3 g^2 \frac{b_0^2}{\kappa(r^2+g^2)^3} \label{eq15}\\ 
     p =-\frac{(3g^{4}-8g^{2}r^{2}+r^{4})b_{0}^{2}}{\kappa(r^{2}+g^{2})^{4}}
=  -\frac{r^{2}(r^{2}-2g^{2})b_{0}^{2}}{\kappa(r^2+g^2)^4} - 3 g^2 \frac{ (g^2-2r^2)b_0^2}{\kappa(r^2+g^2)^4} \label{eq16}
\end{eqnarray}
Notice that the decomposition indicated in the extreme right of Eqns (\ref{eq15}) and (\ref{eq16}) show the parts of $\rho$, $\tau$ and $p$ which are energy condition violating (first term) and energy condition satisfying (second term). Also the second terms in the split shown are exclusively dependent on the regularisation parameter $g$ and reduce to zero when $g=0$. 

\noindent Let us now analyze different classical energy conditions as mentioned above and see whether the `matter required' to support this spacetime violates/satisfies them.\\ 
\begin{figure}[h]
\centering
\includegraphics[width=0.6\textwidth]{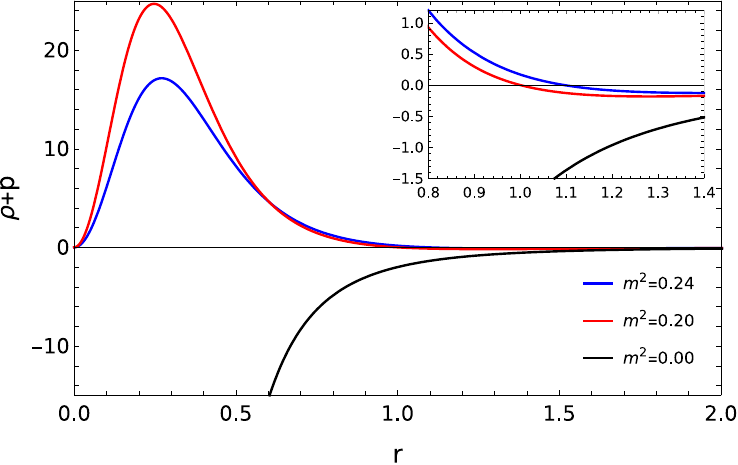}
\caption{Plot of $\rho +p$ with $r$ for different values of $m^{2}$. Blue and red lines represent $\rho +p$ having $m^{2}$ value in the range $0<m^{2}\leq0.25$. The black curve is for the singular metric}
\label{fig:rhowithp}
\end{figure}

\noindent $\bullet$ \textbf{Null Energy Condition :} For all values of $r$, $g^2$, and $b_0^2$, $\rho + p \geq 0$ and $\rho + \tau \geq 0$ must be met in order to satisfy the null energy condition. The first of the criteria in the NEC is met here since, according to equation (\ref{eq15}), $\rho + \tau = 0$. For the second, we have;
\begin{equation}\label{eq17}
    \begin{aligned}
        \rho + p &=\frac{(3g^{2}-r^{2})b_{0}^{2}}{\kappa(r^{2}+g^{2})^{3}}-\frac{(3g^{4}-8g^{2}r^{2}+r^{4})b_{0}^{2}}{\kappa(r^{2}+g^{2})^{4}} \\
        &=\frac{2(r^{2}/b_{0}^{2})(5m^{2}-(r^{2}/b_{0}^{2}))}{\kappa b_{0}^{2}((r^{2}/b_{0}^{2})+m^{2})^{4}}
    \end{aligned}
\end{equation}
\noindent It is clear from the equation above that for all allowed values of $m^2$, $\rho + p \geq 0$ in the range of $r^2\leq 5m^2b_0^2$. As a result, we know exactly what range of $r$ the NEC is fulfilled in. We show how the value of $\rho + p$ varies with $r$ in Figure \ref{fig:rhowithp}. It should be noted that for the singular RN metric $(m^2=0)$  $\rho +p \leq 0$ over the whole domain of $r$. Thus, the regularisation of the singular RN metric$(m^2=0)$ does provide a range of $r$ where the NEC is fulfilled.\\ 
\begin{figure}[h]
\centering
\includegraphics[width=0.6\textwidth]{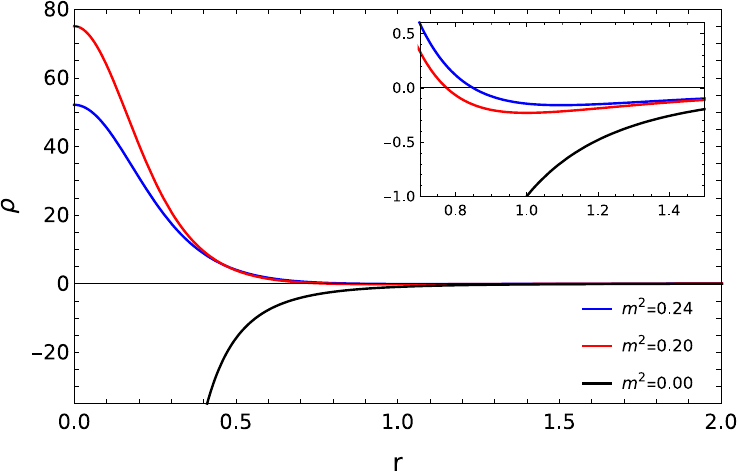}
\caption{Plot of $\rho$ with $r$ for different values of $m^{2}$. Blue and red lines represent $\rho$ having $m^{2}$ value in the range $0<m^{2}\leq0.25$. The black curve is for the singular metric  }
\label{fig:rho}
\end{figure}

\noindent $\bullet$ \textbf{Weak Energy Condition:} The three conditions in the WEC are $\rho \geq 0$, $\rho +\tau \geq 0$, and $\rho+p \geq 0$. According to the analysis for the NEC stated above, the matter required to support our geometry satisfies the second inequality since we have $\rho +\tau =0$. Further, for a given range of $r$,  $\rho +p\geq 0$ also holds, though not everywhere. To analyse the first condition (i.e. $\rho\geq 0$), let us examine eq.(\ref{eq15}). It may be stated as follows:
\begin{equation}\label{eq18}
     \rho =\frac{(3g^{2}-r^{2})b_{0}^{2}}{\kappa (r^{2}+g^{2})^{3}} =\frac{(3m^{2}-(r^{2}/b_{0}^{2}))}{\kappa b_{0}^{2}((r^{2}/b_{0}^{2})+m^{2})^{3}}
\end{equation}
The equation above demonstrates that $\rho\geq0$ for $r^2\leq 3m^2b_0^2$. Figure \ref{fig:rho} illustrates how $\rho$ varies with $r$. From this, it can be inferred that the regularisation of the singular RN geometry provides us with a range of $r$ where $\rho\geq0$ holds.\\
\noindent The end result for WEC reads: $\rho\geq0$ for $r^2\leq 3m^2b_0^2$, $\rho+\tau=0$ for all values of $r$, and $\rho +p\geq0$ in the range of $r^2\leq5m^2b_0^2$. Therefore, it is possible to state that our regular black holes satisfies the WEC for an adjustable finite range of $r$, while the singular RN metric ($m^2=0$) violates WEC over the whole domain of $r$. \\ \\
\noindent $\bullet$ \textbf{Strong Energy Condition:} SEC may be verified by 
additionally looking at $\rho +\tau+2p\geq0$. Since we have $\rho +\tau=0$, studying the behaviour of $p$, one can check SEC.  We may express eq.(\ref{eq16}) as follows:
\begin{equation}\label{eq19}
     p =-\frac{(3g^{4}-8g^{2}r^{2}+r^{4})b_{0}^{2}}{\kappa (r^{2}+g^{2})^{4}}=-\frac{(3m^{4}-8m^{2}(r^{2}/b_{0}^{2})+(r^{2}/b_{0}^{2})^{2})}{\kappa b_{0}^{2}((r^{2}/b_{0}^{2})+m^{2})^{4}}
\end{equation}

\begin{figure}[h]
\centering
\includegraphics[width=0.6\textwidth]{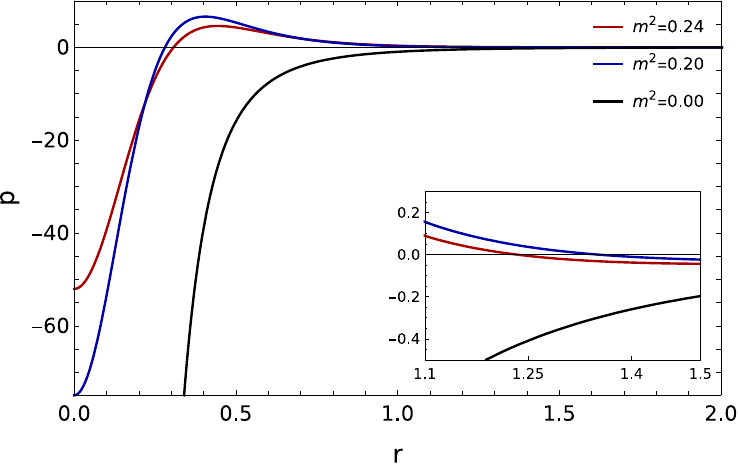}
\caption{Plot of $p$ vs. $r$ for different values of $m^{2}$. Blue and red lines represent $p$ having $m^{2}$ value in the range $0<m^{2}\leq0.25$. The black curve is for the singular metric.}
\label{fig:p}
\end{figure}
\begin{figure}[h]
\includegraphics[width=0.492\textwidth]{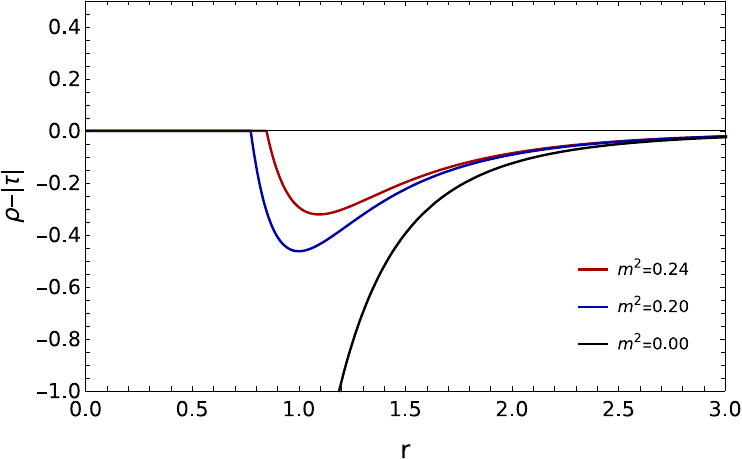}
\includegraphics[width=0.492\textwidth]{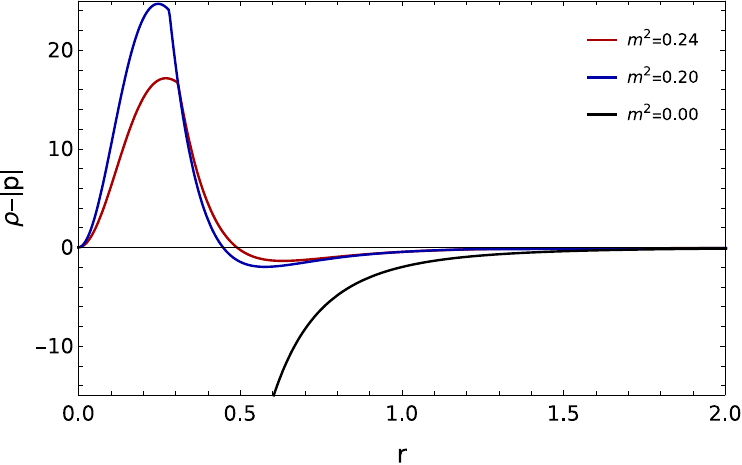}
\caption{Plot of $\rho -|\tau|$ with $r$ for different values of $m^{2}$(left). Variation of $\rho-|p|$ with $r$ for different values of $m^{2}$(right) }
\label{fig:SEC}
\end{figure}

\noindent It is evident from the equation above and by looking at the Figure \ref{fig:p} that SEC is met in the range $(4m^2 - \sqrt{13}m^2 )b_0^2\leq r^2\leq (4m^2 + \sqrt{13}m^2 )b_0^2$. In contrast, the singular RN metric ($m^2=0$) violates the SEC for every $r$. Thus, though not globally, our regular black hole satisfies the SEC at least over a specific range of $r$.
\\ \\
\noindent $\bullet$ \textbf{Dominant Energy Condition:} The DEC comprises of the inequalities $\rho\geq0$, $\rho-|\tau|\geq0$, and $\rho-|p|\geq0$. In the above discussion, we showed that $\rho\geq0$ for $r^{2}\leq3m^{2}b_{0}^{2}$. From eq.(\ref{eq15}-\ref{eq16}) one can write,
\begin{equation} \label{eq20}
    \rho-|\tau| = 
       \begin{cases}
        0, & \text{if } r^{2}\leq 3m^{2}b_{0}^{2}\\
        -\frac{2(r^{2}-3g^{2})b_{0}^{2}}{\kappa (r^{2}+g^{2})^{3}}, & \text{if } r^{2}>3m^{2}b_{0}^{2}
    \end{cases}
\end{equation}
Hence, $\rho-|\tau|\geq0$ is satisfied in the range $r^{2}\leq 3m^{2}b_{0}^{2}$, which is also shown in Figure \ref{fig:SEC}(left). To check the third inequality in DEC, we directly use Figure \ref{fig:SEC}(right). It is easy to see that
\begin{equation}\label{eq21} 
    \rho-|p| = 
       \begin{cases}
        +\text{ve}, & \text{if } r^{2}\leq m^{2}b_{0}^{2}\\
        -\text{ve}, & \text{if } r^{2}>m^{2}b_{0}^{2}
    \end{cases}
\end{equation}
\noindent Therefore, we have $\rho \geq 0$, $\rho-|\tau|\geq 0$ for $r^{2}\leq3m^{2}b_{0}^{2}$ and $\rho-|p|\geq0$ in the range $r^{2}\leq m^{2}b_{0}^{2}$. Hence, the matter for our regular black hole satisfies the DEC in the range $r^{2}\leq m^{2}b_{0}^{2}$.

\noindent Following the above description of energy conditions, the characteristics of the matter needed to maintain our regular black hole may be summed up as follows:

\begin{center}
\begin{tabular}{ |c|c| } 
\hline
Energy condition & Range of validation  \\
\hline
NEC & $r^{2}\leq5m^{2}b_{0}^{2}$ \\ 
WEC & $r^{2}\leq3m^{2}b_{0}^{2}$ \\
SEC & $(4-\sqrt{13})m^{2}b_{0}^{2}\leq r^{2}\leq (4+\sqrt{13})m^{2}b_{0}^{2}$\\
DEC & $r^{2}\leq m^{2}b_{0}^{2}$\\
\hline
\end{tabular}
\end{center}
As a result, we may infer that our regular black hole solution fulfils all the energy conditions known in classical GR, in the range $(4-\sqrt{13})m^2b_0^2\leq r^2\leq m^2b_0^2$. However, the singular RN type metric $(m^2=0)$ violates each and every energy condition across the whole range of $r$. 

\section{Matter sources for our geometries}\label{III}
\noindent We now turn to the question of constructing sources that may be used to model the matter stress-energy required to support the spacetimes mentioned above. Sources for regular black hole spacetimes have been constructed using nonlinear electrodynamics, scalar fields, and phantom scalar fields \cite{Ayon1, Ayon2, Ayon3, Ayon4, Ayon5, Bronikov2, Bronnikov5, Bronnikov6}. Though some of the sources may be physically questionable, they do provide some decent examples of Lagrangian-based matter sources for regular geometries. Let us now try and see what could be the possible sources for 
the spacetimes proposed in this article. The necessity of `exotic matter fields' is evident from the analysis of energy conditions studied in the earlier section. The well-known Bardeen and Hayward regular black holes have been interpreted as gravitational fields generated by a nonlinear magnetic monopole, as constructed in \cite{Ayon4, Fan}. However, the Lagrangian describing the dynamics of the fields $(\mathbf{F})$ are different in the above two cases, i.e. we don't have a single well-defined Lagrangian for both. A more general matter Lagrangian for Bardeen and Hayward spacetimes (as well as others) is presented in \cite{Zhong}. It is given as 
\begin{equation}\label{eq22}
    \mathcal{L(\mathbf{F})}=\frac{4\mu}{\delta}\frac{(\delta \, \mathbf{F})^{\frac{\nu+3}{4}}}{(1+(\delta \, \mathbf{F})^{\frac{\nu}{4}})^{\frac{\mu+\nu}{\nu}}}
\end{equation}
where $\mu>0$, $\nu>0$ are dimensionless constants and $\delta>0$ has the dimension of length squared. One obtains a source for the Bardeen metric for $\mu=3, \nu=2$ and for the Hayward metric for $\mu=\nu=3$.  The matter sources arising from the above Lagrangian satisfy the null energy condition and violate the strong energy condition. Though, in our case, we have a violation of the null energy condition, our spacetime geometry, as well as the well--known regular black holes satisfy a common relation for the required stress energy, namely $\rho + \tau =0$. In addition, we also have the equality of tangential pressures arising out of spherical symmetry. Below we have depicted two probable sources for our geometries: a non-linear electrodynamic field and another using ideas from braneworld gravity. 

\subsection{Nonlinear electrodynamics}
\noindent Motivated by the above discussion, we consider a non-linear electromagnetic field Lagrangian as a source for our regular black hole metric. We show in eq.(\ref{eq15}-\ref{eq16}) how one can decompose the diagonal elements of the stress-energy tensor in two parts -- one satisfying the NEC,  WEC, and the other violating them. Interestingly, both parts of the decomposition satisfy $\rho +\tau=0$ separately. One may therefore propose a non-linear electromagnetic field Lagrangian $\mathcal{L}=\mathcal{L}_{1}(\mathbf{F})+\mathcal{L}_{2}(\mathbf{F})$  minimally coupled to gravity by the following action,
\begin{equation}\label{eq23}
    S=\int d^{4}x \sqrt{-g}\left(\frac{R}{\kappa}-\mathcal{L}_{1}(\mathbf{F})-\mathcal{L}_{2}(\mathbf{F})\right),
\end{equation}
where $R$ is the Ricci scalar. $\mathcal{L}_{1}(\mathbf{F})$ is the non-linear electromagnetic Lagrangian which generates an energy--momentum tensor violating the NEC, WEC. $\mathcal{L}_{2}(\mathbf{F})$, in contrast, leads to a ${T_{\mu\nu}}$ which satisfies the WEC, NEC. Here, $F_{\mu \nu}$ is electromagnetic strength tensor and $\mathbf{F}\equiv F^{\mu \nu}F_{\mu \nu}$.

\noindent The covariant equations of motion (field variation and metric variation) which emerge from the above Lagrangian are given as,
\begin{align}\label{eq24}
    \nabla_{\mu}\left(\frac{\partial\mathcal{L}_{(i)}}{\partial\mathbf{F}}F^{\mu \nu}\right)=0; \hspace{0.4cm} i=1,2\\
    R_{\mu \nu}-\frac{1}{2}g_{\mu \nu}R=\kappa \mathbf{T}_{\mu \nu}=\kappa \sum_{i=1}^{2}T_{\mu \nu}^{(i)}
\end{align}
where $R_{\mu \nu}$ is the Ricci tensor. $T_{\mu \nu}^{(1)}$ and $T_{\mu \nu }^{(2)}$ are energy-momentum tensors corresponding to $\mathcal{L}_{1}(\mathbf{F})$ and $\mathcal{L}_{2}(\mathbf{F})$, respectively. One can write the two energy-momentum tensors in the following fashion,
\begin{align}\label{eq26}
    T_{\mu \nu}^{(i)}=2\left(\frac{\partial\mathcal{L}_{(i)}}{\partial\mathbf{F}} F_{\mu \alpha}F_{\nu}{}^{\alpha}-\frac{1}{4}g_{\mu \nu} \mathcal{L}_{(i)}(\mathbf{F})\right); \hspace{0.2cm} i=1,2
\end{align}
From eq.(\ref{eq15}-\ref{eq16}), it is clear that we need a matter source that has isotropic tangential pressures and satisfies $\rho+\tau=0$. As a result, we only have non-zero components of the electromagnetic strength tensor $F_{tr}$ and $F_{\theta \phi}$ to serve this purpose.

\noindent Hence, in the frame basis, the diagonal elements of the total energy-momentum tensor take the following form:
\begin{align}\label{eq27}
    \mathbf{T}_{00}=&-\mathbf{T}_{11}=\sum_{i=1}^{2}\left(\frac{1}{2}\mathcal{L}_{(i)}(\mathbf{F})-2\frac{\partial\mathcal{L}_{(i)}}{\partial\mathbf{F}}F_{tr}F^{tr}\right);\\
    \mathbf{T}_{22}=&\mathbf{T}_{33}=\sum_{i=1}^{2}\left(-\frac{1}{2}\mathcal{L}_{(i)}(\mathbf{F})+2\frac{\partial\mathcal{L}_{(i)}}{\partial\mathbf{F}}F_{\theta\phi}F^{\theta\phi} \right)
\end{align}
It is evident from the above that for any ${\cal L} (\mathbf {F})$, one does have $\rho+\tau=0$ and isotropic tangential pressures. From the Einstein tensor components evaluated earlier in the frame basis and noticing the split as shown in eq.(\ref{eq15}-\ref{eq16}), one can write
\begin{align}
    \frac{1}{2}\mathcal{L}_{1}(\mathbf{F_0})-2F_{tr}F^{tr}\left ( \frac{\partial\mathcal{L}_{1}}{\partial\mathbf{F}} \right )_{\mathbf{F}=\mathbf{F}_{0}}=&-\frac{b_0^2 r^2}{\kappa(r^2+g^2)^3};  \label{eq29}\\
    -\frac{1}{2}\mathcal{L}_{1}(\mathbf{F_0})+2F_{\theta \phi}F^{\theta \phi} \left ( \frac{\partial\mathcal{L}_{1}}{\partial\mathbf{F}} \right )_{\mathbf{F}=\mathbf{F_0}}=&-\frac{b_0^2 r^2(r^2-2g^2)}{\kappa(r^2+g^2)^4}; \label{eq30}\\
    \frac{1}{2}\mathcal{L}_{2}(\mathbf{F_0})-2F_{tr}F^{tr}\left ( \frac{\partial\mathcal{L}_{2}}{\partial\mathbf{F}} \right )_{\mathbf{F}=\mathbf{F}_{0}}=&+\frac{3 g^{2}b_0^2}{\kappa(r^2+g^2)^3}; \label{eq31}\\
    -\frac{1}{2}\mathcal{L}_{2}(\mathbf{F_0})+2F_{\theta\phi}F^{\theta\phi}\left ( \frac{\partial\mathcal{L}_{2}}{\partial\mathbf{F}} \right )_{\mathbf{F}=\mathbf{F_0}} =&- 3 g^2 \frac{b_0^2 (g^2-2r^2)}{\kappa (r^2+g^2)^4};\label{eq32}
\end{align}
where the LHS in the above equations are evaluated on-shell, i.e. for a specific solution $\mathbf{F}= \mathbf {F_0}$, which, of course, has to satisfy the field equations.

\noindent Let us assume that for specific values of parameters, a combination of the following two non-linear electrodynamic field Lagrangian densities may support our regular black holes;
\begin{align}\label{eq33}
    \mathcal{L}_{1}(\mathbf{F})=-\frac{\gamma \mathbf{F}}{\left(1+\eta \sqrt{\mathbf{F}}\right)^{3}}; \hspace{0.8cm}\mathcal{L}_{2}(\mathbf{F})=\frac{3\gamma \eta \mathbf{F}^{3/2}}{\Big(1+\eta \sqrt{\mathbf{F}}\Big)^{3}}
\end{align}
where $\gamma$ and $\eta$ are two separate constants. Therefore, one might have distinct field dynamics based on the different $\gamma$ and $\eta$ options. For our purposes, we select just a magnetic solution with $\mathbf{F_0}=2F_{\theta \phi }F^{\theta \phi}=\frac{2q_{m}^{2}}{r^4}$, $\gamma=b_0^2/\kappa q_{m}^2$, and $\eta=g^2/\sqrt{2q_{m}^2}$, which satisfy eq.(\ref{eq29}-\ref{eq32}), where $q_{m}^2$ is an arbitrarily chosen constant. The metric 
parameters $b_0$ and $g$ are both related to the `magnetic charge $q_m$',
as is apparent from the expressions above.

\noindent Therefore, the total non-linear electrodynamic field Lagrangian density can be written as a combination of the above two in the following way;
\begin{equation}\label{eq34}
    \mathcal{L}(\mathbf{F})=-\frac{\gamma\mathbf{F}(1-3\eta\sqrt{\mathbf{F}})}{\Big(1+\eta\sqrt{\mathbf{F}}\Big)^{3}}
\end{equation}
Note that for $g^{2}=0$ or $\eta=0$, i.e. singular RN type solution, the matter source becomes;
\begin{equation}\label{eq35}
    \mathcal{L}(\mathbf{F})=-\gamma \mathbf{F}
\end{equation}
Thus, the singular RN-type solution with `imaginary' charge is sourced by a linear magnetic Lagrangian density. The source Lagrangian of equation (\ref{eq34}) has the limitation that it cannot allow negative $\mathbf{F}$ values since it contains fractional powers of $\mathbf{F}$. Therefore, we are limited to working with positive $\mathbf{F}$ values. Hence, we
can have configurations with both electric and magnetic fields or only a magnetic field. But any scenario with only
an electric field is excluded. However, it must be noted that the total Lagrangian,
when expanded in $\eta$ has `corrections' (beyond the Maxwell term) 
involving $\mathbf{F} \sqrt{\mathbf {F}}$ and $\mathbf{F^2}$ at orders $\eta$ and $\eta^2$. Since $\eta$ is proportional to $g^2$, it is only in the gravitational field of the regular black hole do these terms acquire a meaning. The absence of a purely electric configuration (as implied through the
presence of a square root of $\mathbf{F}$)
is a feature (deficiency?)  not specific to our solution alone, but is generically found
in some form or the other in all source models for regular black holes using nonlinear electrodynamics.\cite{Ayon1,Ayon3,Ayon2,Ayon4,Ayon5}

\noindent  As stated earlier, our regular black holes are made up of the two parameters $b_0^2$ and $g^2$. However, the NED source Lagrangian depends on an additional free parameter, $q_{m}^2$. Thus, to understand what $q_{m}^2$ is and the behaviour of the source current that creates this type of magnetic field, we must investigate the dynamics of the source Lagrangian. 

\noindent The functional value of $\mathbf{F_0}$ makes it clear that $F_{\theta\phi}=-q_{m}\sin{\theta}$.  Further, in curved spacetime, the non-linear Maxwell-like equations which follow from the Lagrangians mentioned above, with source current, may also be expressed as,
\begin{align}\label{eq36}
    \nabla_{\mu}F^{\mu \nu}+\left(\frac{\partial^{2}\mathcal{L}_{(i)}}{\partial\mathbf{F}^{2}}\bigg/\frac{\partial\mathcal{L}_{(i)}}{\partial\mathbf{F}}\right)(\partial_{\mu}\mathbf{F})F^{\mu \nu}=J^{\nu}_{e}{}_{(i)}\\
    \nabla_{\mu}\Tilde{F}^{\mu \nu}=J^{\nu}_{m}
\end{align}
where $\Tilde{F}^{\mu \nu}\equiv\frac{1}{2}\epsilon^{\mu \nu \rho \sigma}F_{\rho \sigma}$ ($\epsilon^{\mu \nu \rho \sigma}$ is the Levi-Civita tensor), the dual of field strength tensor. For the derived field strength tensor $F_{\theta\phi}$, we find that $J^{\nu}_{e}=\{0,0,0,0\}$ and $J^{\nu}_{m}=\{\rho_{m},0,0,0\}$ (where, $\rho_{m}=q_{m}\delta^{3}(r)$) independent of the form of source Lagrangian. Therefore, it is also valid for the total electrodynamic field Lagrangian $(\mathcal{L}(\mathbf{F}))$. Thus, a non-linear magnetic monopole at $r=0$ supports our regular black hole, with $q_{m}$ being the total magnetic charge.

\subsection{Braneworld gravity}
\begin{figure}[h]
         \includegraphics[width=0.492\textwidth]{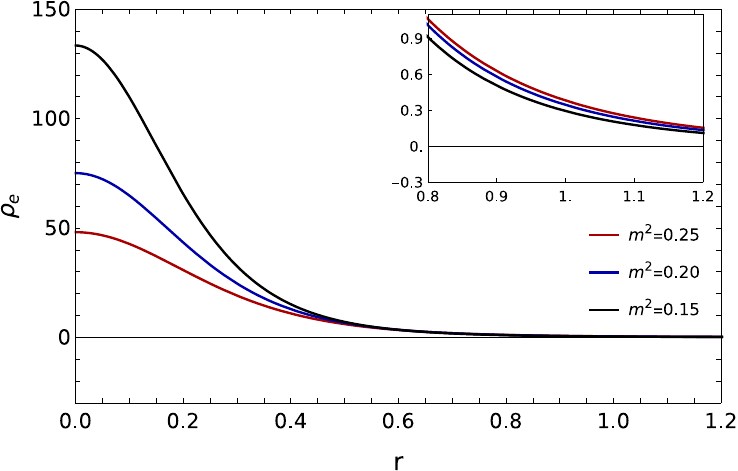}
         \includegraphics[width=0.492\textwidth]{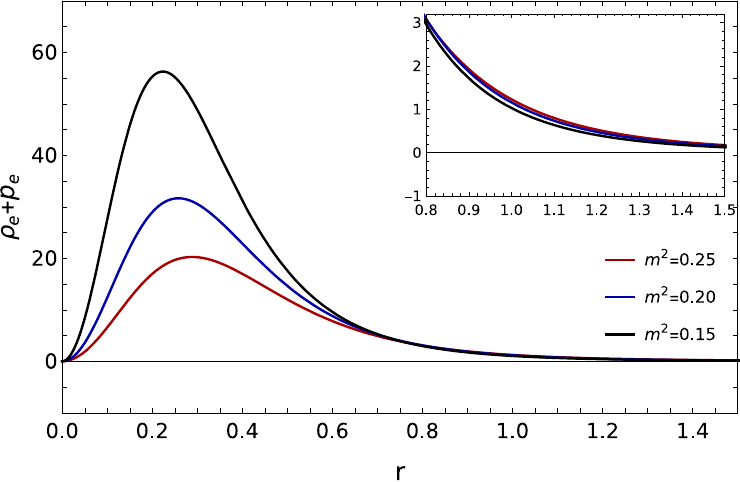}
        \caption{(Left) Plot of $\rho_{e}$ with $r$ shows a positive definite energy density of the extra matter part. (Right) Plot of $\rho_{e}+p_{e}$ with $r$, which is also positive for all values of $r$. }
        \label{fig:extramatter}
\end{figure}

\noindent As stated earlier, we may also consider our regular black hole geometry in the context of {\em brane-world gravity} \cite{Maartens}. To facilitate this, let us introduce a new kind of decomposition of the energy-momentum tensor components in eq.(\ref{eq15}-\ref{eq16}).
\begin{align}
    \rho=-\tau=-\frac{b_{0}^{2}r^{2}}{\kappa(r^{2}+g^{2})^{3}}+\frac{3g^{2}b_{0}^{2}}{\kappa(r^{2}+g^{2})^{3}}=\rho_{t}+\rho_{e}\label{eq38}\\
    p=-\frac{b_{0}^{2}r^{2}}{\kappa(r^{2}+g^{2})^{3}}-3g^{2}b_{0}^{2}\frac{(g^{2}-3r^{2})}{\kappa(r^{2}+g^{2})^{4}}=p_{t}+p_{e}\label{eq39}
\end{align}
One may note that we have chosen a decomposition such that a part of the energy-momentum tensor is traceless. The rest of the energy-momentum is non-zero only when $g\neq 0$ and it satisfies the NEC, WEC, as is evident from the Figure \ref{fig:extramatter}. We can exploit this decomposition as follows. Recall the well-known effective Einstein equations on the four-dimensional 3-brane embedded in an
ambient, curved five dimensional spacetime. They are written as,
\begin{equation}\label{eq40}
    G_{\mu \nu}=-\Lambda g_{\mu \nu} +\kappa T_{\mu \nu} +6\frac{\kappa}{\lambda}S_{\mu \nu}-\mathcal{E}_{\mu \nu}
\end{equation}
where $\Lambda$ is a four dimensional cosmological constant on the 3-brane, $T_{\mu\nu}$ is the on-brane matter, $S_{\mu\nu}$ is the so-called quadratic stress energy on the brane. The traceless $\mathcal{E}_{\mu \nu}$ is a geometric quantity controlled by the extra dimension and related to the Weyl tensor of the bulk five-dimensional spacetime. If we assume a large brane tension $\lambda$ and a zero $\Lambda$, the terms involving $S_{\mu\nu}$ drop out, and we are left with only the $\mathcal{E}_{\mu \nu}$, which can be used to model the traceless part in the expressions in (\ref{eq38}), (\ref{eq39}). Note that the negativity of $\rho_t$ and $p_t$ is already accounted for through the negative sign in front of the $\mathcal{E}_{\mu \nu}$ term in the effective Einstein equations in (\ref{eq40}). The remaining part, i.e. the $\rho_e$ and $p_e$ can be adjusted into the `on-brane matter' described through $T_{\mu\nu}$.

\noindent As is known, the RN spacetime with negative $Q^2$ (here
$b_0^2$) was used as one of the first solutions of the effective Einstein equations on the brane (see \cite{Maartens} and references therein). The $b_0^2$ in this context is related to a
`tidal charge parameter' induced on the brane from the bulk
through ${\cal E}_{\mu\nu}$. Regularisation of this geometry seems to add extra on-brane matter, which is well-behaved insofar as energy conditions are concerned. 

\section{Shadow radius and observational bounds}\label{IV}

\noindent Finally, we ask the question--does the above family of solutions have any
relevance w.r.t. observations? One could calculate various quantities--such as 
light deflection, orbits, time delay etc, while looking for possible signatures.
We choose to work out the shadow profiles since the recent EHT results 
may then be used to provide bounds on the metric parameters.

\noindent The strong gravitational field around a black hole causes photons to be deflected or absorbed, which results in a silhouette which is also termed as the black hole shadow. Each black hole has a unique shape and size for its shadow, which is controlled by parameters appearing in its geometry (i.e. metric parameters such as mass, charge, angular momentum etc.). Originally, Bardeen \cite{bardeenshadow} introduced (assuming the observer very far away) celestial coordinates in the observer's sky to describe and quantify the shadow. More recently, there have been attempts wherein shadows have been calculated using ray tracing methods for observers (moving or static) at a finite distance or in asymptotically non-flat scenarios \cite{Volker}.

\noindent The shadow of a black hole may be seen as a dark region surrounded by photon rings.  Photons with less angular momentum scatter from the black hole and may be seen from infinity; photons with higher angular momentum enter the black hole and leave a dark circle. As mentioned above, two celestial coordinates, $\alpha$ and $\beta$, were introduced by Bardeen which are,
\begin{equation}\label{eq41}
    \alpha=\lim_{r_{0}\to\ \infty}\Big(-r_{0}^{2}\sin{\theta_{0}}\frac{d\phi}{dr}\Big); \hspace{1cm} \beta=\lim_{r_{0}\to\ \infty}r_{0}^{2}\frac{d\theta}{dr}
\end{equation}
where $r_{0}$ is the distance between the black hole and the observer, $\theta_{0}$ is the inclination angle between the black hole rotation axis and the line of sight between the source and observer. The derivatives $\frac{d\theta}{dr}$ and 
$\frac{d\phi}{dr}$ have to be evaluated in the asymptotic region using the
first integrals of the geodesic equations. The radius of the static spherically symmetric black hole shadow, $r_{sh}$, as viewed by a static observer at radial coordinate $r_{0}$, may be approximated as \cite{Volker}:
\begin{equation}\label{eq42}
    r_{sh}^{2}=\alpha^{2}+\beta^{2}=\frac{r_{ph}^{2}}{w(r_{ph})}
\end{equation}
where $r_{ph}$ is the radius of photon sphere and $w(r)=-g_{tt}(r)$. For a static spherically symmetric black hole, the photon sphere is the orbit at which light moves in an unstable circular null geodesic \cite{Virbhadra}. The equations of motion for photons around a static and spherically symmetric black hole environment can be obtained using the Hamilton-Jacobi or Hamiltonian formulations, as described in \cite{Chen}, which then enables one to determine the effective potential defining the system. The radius of the photon sphere $r_{ph}$ can be calculated from the critical point conditions of this effective potential by solving the expression
\begin{equation}\label{eq43}
    \frac{r_{ph}w^{'}(r_{ph})}{w(r_{ph})}=2
\end{equation}
where a prime denotes a derivative with respect to the radial coordinate.

\subsection{Shadow radius of our regular black hole}

\noindent Based on the description above, we first need to know the radius of the photon sphere in order to calculate the shadow radius of our black hole. The equation for the radius of the photon sphere can be obtained by substituting $w(r)=1-\frac{b_0^2r^2}{(r^2+g^2)^2}$ in equation (\ref{eq43}), which gives,
\begin{equation}\label{eq44}
    (r_{ph}^{2}+m^{2}b_{0}^{2})^{3}-2r_{ph}^{4}b_{0}^{2}=0
\end{equation}
where we use our special parametrization $g^{2}=m^{2}b_{0}^{2}$. When
$m=0$, i.e. $g=0$ and $b_0\neq 0$ (the singular geometry) 
the photon sphere is at $r_{ph}= {\sqrt 2}b_0$. For $m\neq 0$,
the above expression represents a cubic algebraic equation in the variable $r_{ph}^2$. To analyse the characteristics of the roots, it is convenient to express the cubic equation in $r_{ph}^2$ in the {\em depressed cubic form} and examine its discriminant. We have checked (not shown here) through plots
that the discriminant of the cubic equation turns out to be positive or equal to zero for values of $m^2$ within the range of $0$ to $0.296$. This observation provides confirmation that there are three distinct real roots for $r_{ph}^2$ within the specified range of $m^2$. Hence, within the specified interval of $m^2$, it is possible to obtain two positive real roots, two negative real roots, and two imaginary roots for $r_{ph}$. We take into consideration the two real positive roots as potential candidates for the photon sphere since they lie inside the domain of coordinates. An unstable circular orbit may now be found at one of these two roots by performing a stability analysis on the effective potential, which is defined as,
\begin{equation}\label{eq45}
    V_{eff}(r)=\Big(1-\frac{b_{0}^{2}r^{2}}{(r^{2}+m^{2}b_{0}^{2})^{2}}\Big)\frac{L^{2}}{r^{2}}
\end{equation}
Here $L$ is the angular momentum of the photon. Let us say $\Bar{r}_{ph}$ is the location of the unstable circular orbit. Hence, the shadow radius from eq.(\ref{eq42}) becomes;
\begin{equation}\label{eq46}
r_{sh}^{2}=\frac{\Bar{r}_{ph}^{2}(\Bar{r}_{ph}^{2}+m^{2}b_{0}^{2})^{2}}{(\Bar{r}_{ph}^{2}+m^{2}b_{0}^{2})^{2}-\Bar{r}_{ph}^{2}b_{0}^{2}}
\end{equation}
 If $m=0$ (i.e. $g=0$, $b_0\neq 0$), the shadow radius turns out to be
given as $r_{sh} = 2 b_0$ (on using the photon sphere radius $\Bar{r}_{ph}=\sqrt{2}b_0$).
Recall that in section \ref{second}, it has been demonstrated that a horizon exists for values of $m^2$ such that $0<m^2\leq0.25$. In this specified interval of $m^2$, we find that $\Bar{r}_{ph}$ can vary between $1.141 b_{0}> \Bar{r}_{ph}\geq1.029 b_{0}$. However, for $0.25<m^{2}\leq0.296$, there is no horizon but a photon sphere exists, i.e. we have a regular spacetime having a photon sphere which may be used to model a compact object.  In addition, the asymptotic observer can estimate the angular diameter of the shadow as: $2\alpha_{D}=2r_{sh}/r_{0}$ \cite{Volker}.

\subsection{Constraints from the observed shadow of M87* and Sgr A*}

\noindent Even though astrophysical black holes rotate and the shape of the shadow is also controlled by the rotation parameter, one can still use the simple circular shadow of a static and spherically symmetric black hole (without rotation) to make a preliminary, qualitative
estimate of the black hole parameters. In addition, one recalls
that the observed shadows for M87$^{*}$ and Sgr A$^{*}$ are reported to be quite close to being circular, with any deviation from circularity
being small. This motivates us to identify and check possible observational traces of our regular black hole in the supermassive compact objects in M87$^*$ and Sgr A$^*$. We will use freely available EHT results in order to arrive at our estimates. 

\noindent In this section, we have estimated the metric parameter $b_{0}$ for different $m^{2}$ values using the shadows observed by EHT. It is important to keep in mind that the theoretically obtained angular diameter depends on the black hole parameters $b_{0}$, $m^{2}$, and the black hole to observer's distance, $r_{0}$. Thus, to estimate the parameter $b_0$ from the observed angular diameter data, one requires an independent measurement of the distance $(r_0)$ and a hand-picked value of $m^2$.

\begin{figure}[h]
\includegraphics[width=0.492\textwidth]{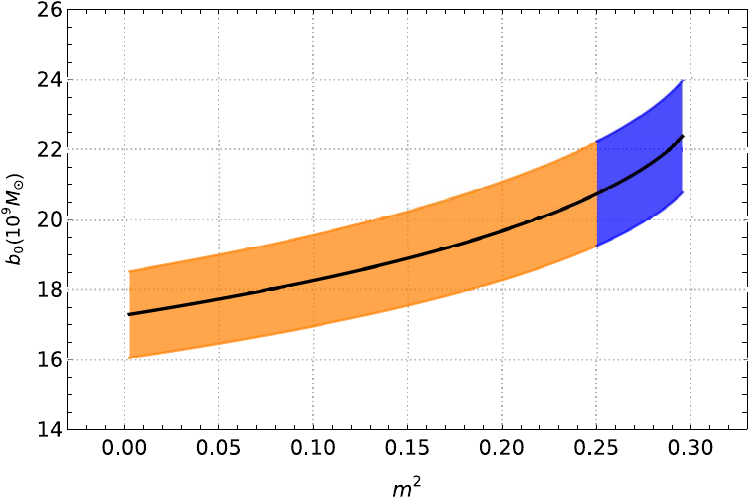}
\includegraphics[width=0.492\textwidth]{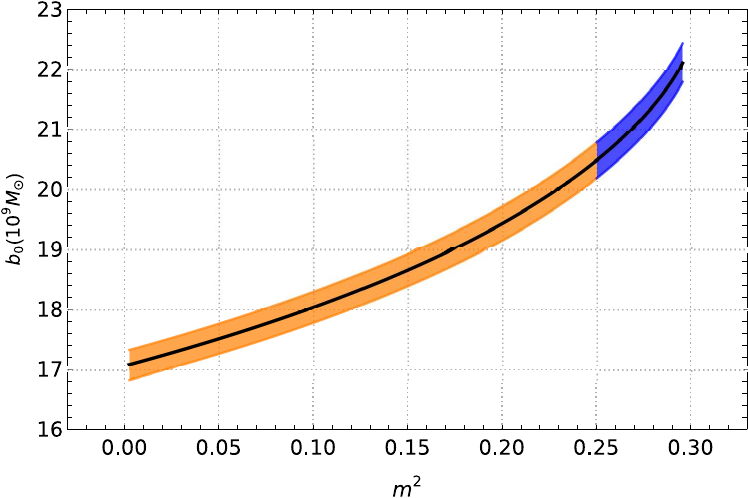}     
\caption{The above figures demonstrate the variation of $b_{0}$ with $m^{2}$ for the observed angular diameter of M87$^{*}$, considering both analyses of shadow diameter reported as $42\pm 3\mu as$ (left) and $41.5\pm0.6\mu as$ (right)  }
\label{fig:M87}
\end{figure}

\noindent In the 2019 EHT reports, the observed angular diameter of black hole M87$^{*}$ at the centre of the galaxy M87 is $\Phi=(42\pm 3)\mu as$ \cite{Akiyama1, Akiyama2, Akiyama3, Akiyama4, Akiyama5, Akiyama6}. Recently, the original  EHT data was more thoroughly analysed in \cite{Medeiros}, and they obtained $\Phi=41.5\pm 0.6\mu as$. We estimate our metric parameters using both results. According to {\em stellar population measurement}, M87$^*$ is located at a distance of $r_0=(16.8\pm 0.8) Mpc$ \cite{Blakeslee,Bird, Cantiello}. However, the error bar of the distance measurements has been disregarded for our purposes. Using earlier formulae, the following expression provides the theoretical angular diameter of the shadow of M87$^*$ provided it is modeled by our regular spacetime geometry:
\begin{equation}\label{eq47}
    2\alpha_{D}=8189.26\times10^{-16} \sqrt{\frac{\Bar{r}_{ph}^{2}(\Bar{r}_{ph}^{2}+m^{2}b_{0}^{2})^{2}}{(\Bar{r}_{ph}^{2}+m^{2}b_{0}^{2})^{2}-\Bar{r}_{ph}^{2}b_{0}^{2}}} \hspace{0.15in}\mu as
\end{equation}
\noindent where we have substituted $r_{0}=16.8 Mpc$. Now, one may calculate $\Bar{r}_{ph}$ for a given value of $m^2$ using the photon sphere equation (\ref{eq44}) and by the stability analysis of the effective potential. Substituting $\Bar{r}_{ph}$ in equation (\ref{eq47}), we have the theoretical angular diameter of the shadow as a function of $b_{0}$. Thus, one may estimate $b_{0}$ by comparing this theoretical prediction with observed data. In Figure [\ref{fig:M87}], we have shown a 2D plot of $(m^{2},b_{0})$ values. The shaded region contains permissible values of $(m^2,b_0)$ based on the observed angular diameter of the shadow of M87$^*$ with its associated uncertainty, as per the observed data reported in \cite{Akiyama1, Akiyama2, Akiyama3, Akiyama4, Akiyama5, Akiyama6, Medeiros}.
As mentioned above, we have shown two plots corresponding to the two data analyses of the same EHT data of M87$^{*}$. The central black lines represent the set of values of $(m^{2},b_{0})$ corresponding to the 
shadow angular diameter value of $42 \mu as$ (Fig. \ref{fig:M87}(left)) and $41.5 \mu as$ (Fig. \ref{fig:M87}(right)), respectively. The error in the measurement of the angular diameter is also considered and has been shown with orange and blue shaded regions in the parameter space. For the angular diameter $(42\pm 3)\mu as$, in Figure \ref{fig:M87}(left), the orange-shaded region corresponds to the regular black hole $(0<m^{2}\leq 0.25)$ having the value of metric parameter $b_{0}$ between $16.05\times10^{9}M_\odot<b_{0}\leq 22.22\times 10^{9}M_\odot$ and the blue shaded region depicts the values of $(m^2,b_0)$ which could represent a horizonless compact object $(0.25<m^{2}\leq0.296)$. Assuming instead, the angular diameter value as $41.5\pm0.6\mu as$, in Figure \ref{fig:M87}(right), the value of $b_{0}$ for the regular black hole varies between $16.83\times10^{9}M_\odot<b_{0}<20.78\times10^{9}M_\odot$. The horizonless compact object can also model the same data, as represented in the blue shaded region of Figure \ref{fig:M87}(right).  
\begin{figure}[h]
         \includegraphics[width=0.492\textwidth]{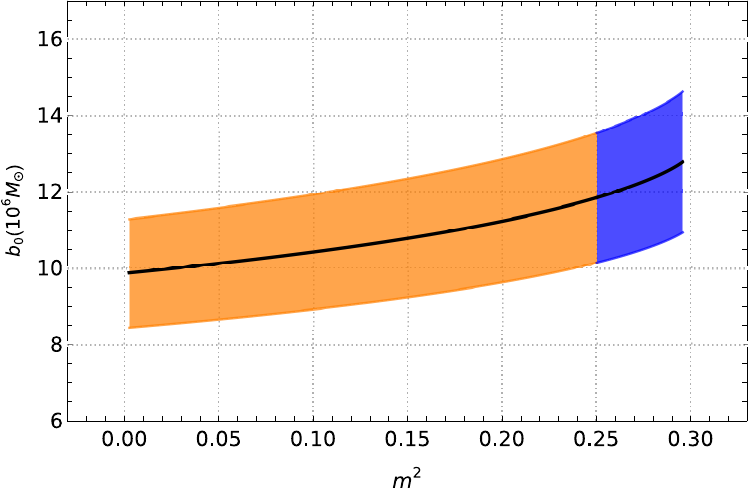}
         \includegraphics[width=0.492\textwidth]{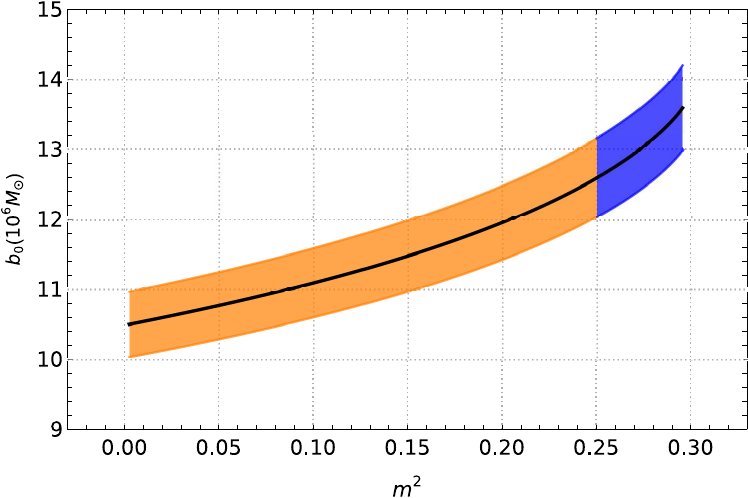}
         \includegraphics[width=0.492\textwidth]{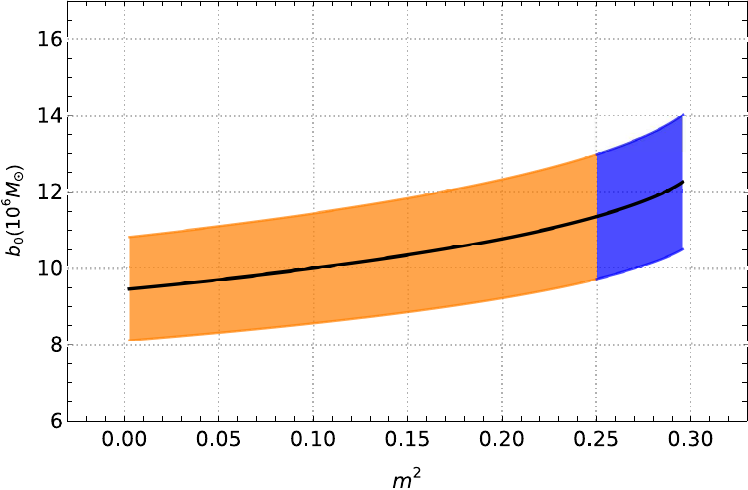}
         \includegraphics[width=0.492\textwidth]{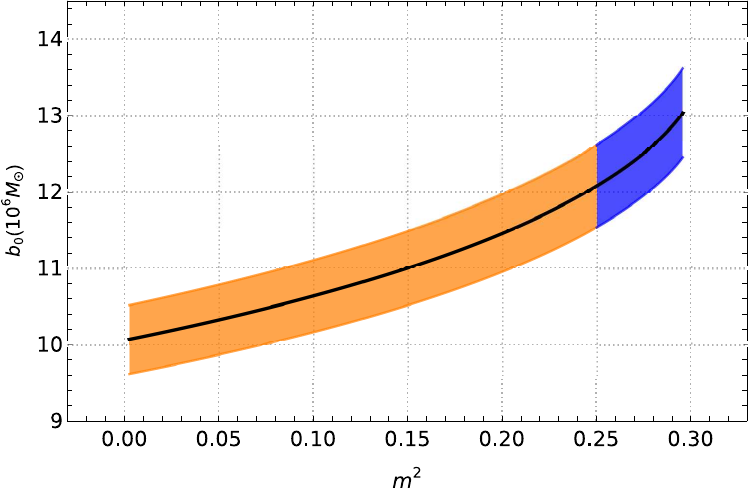}
        \caption{Above figures illustrate the variation of $b_{0}$ with $m^{2}$ for angular diameter of  Sgr A$^{*}$, considering both shadow diameter $48.7\pm 7\mu as$ (top left and bottom left) and emission ring diameter $51.8\pm 2.3\mu as$ (top right and bottom right), while distances are taken to be those reported by the Gravity collaboration (top panel) and Keck team (bottom panel).}
        \label{fig:SgrAG}
\end{figure}

\noindent One can perform a similar exercise with the observed shadow of Sgr A$^{*}$. According to EHT collaboration the angular diameter of the shadow for Sgr A$^{*}$ is $(48.7\pm 7) \mu as$  and emission ring of Sgr $A^{*}$ has angular diameter $51.8\pm 2.3 \mu as$ \cite{Akiyama7, Akiyama8, Akiyama9, Akiyama10, Akiyama11, Akiyama12}. There are several works on the shadow of Sgr A$^{*}$ using either shadow diameter ($48.7\pm 7 \mu as$ ) or emission ring diameter ($51.8\pm 2.3 \mu as$). Therefore, we estimate the metric parameter $b_{0}$ by analyzing both data sets, for completeness. There are also reports on the distance of Sgr A$^{*}$. Keeping the redshift parameter-free, the {\em Keck team} reported the distance of Sgr A$^{*}$ to be $(7959\pm59\pm32)pc$ \cite{Do}. Considering the redshift parameter unity, the same group reported the distance as $(7935\pm50)pc$ \cite{Do}. According to the {\em Gravity collaboration}, the distance is $(8246.7\pm9.3)pc$ \cite{Gravity1, Gravity2}. By accounting for optical aberrations, the {\em Gravity Collaboration} significantly reduced the BH distance to $(8277\pm 9 \pm33)pc$.
Figures \ref{fig:SgrAG} (top panel) and \ref{fig:SgrAG} (bottom panel) represent the parameter space $(m^2,b_0)$ corresponding to the observer to Sgr A$^*$ distance as reported by {\em Gravity collaboration} and {\em Keck team}, respectively. Each panel of Figure \ref{fig:SgrAG}(top, bottom) contains separate analyses of the parameter space by assuming both shadow diameter and emission ring diameter of Sgr A$^*$. Thus plots of Figure \ref{fig:SgrAG} demonstrate how both horizonless compact objects and regular black holes may model the shadow of Sgr A$^*$. 

\noindent As a result, we obtain a collection of points $(m^2,b_0)$ that can be used to simulate the angular diameter of the shadows of M87$^*$ and Sgr A$^*$. These points represent either a regular black hole or a horizonless compact object. In Table \ref{table1}, we have summarized the limits on $b_0$ imposed by the shadows of M87$^*$ and Sgr A$^*$. 

\newcolumntype{M}[1]{>{\centering\arraybackslash}m{#1}}
\newcolumntype{N}{@{}m{0pt}@{}}
\begin{table}[h]
\centering
\begin{tabular}{ |M{1.5cm}|M{2.5cm}|M{2.6cm}|M{3.5cm}|M{3.5cm}|M{1.5cm}|N } 
\hline
Massive Object& Distance from the observer  & Angular Diameter$(\mu as)$ & Constraint on $b_0$ for regular BH & Constraint on $b_0$ for compact object& Unit of $b_0$   \\[5pt]
\hline
\multirow{2}{*}{M87$^*$} & \multirow{2}{2.5cm}{$16.8$ Mpc (Stel. popula. measure.)} &$42\pm 3$  & $16.05<b_0\leq 22.22$ & $19.26<b_0\leq24.01$& \multirow{2}{*}{$10^{9} M_\odot$} \\[10pt] 
\cline{3-5}
 & & $41.5\pm 0.6$ & $16.83<b_0\leq 20.78$ & $20.19<b_0\leq 22.46$& \\[10pt] 
 \hline
\multirow{4}{*}{Sgr A$^*$ }& \multirow{2}{2.5cm} {$8277$ pc (Gravity Collab.)} & $48.7\pm 7$ & $8.46<b_0\leq 13.55$ & $10.14<b_0\leq 14.64$& \multirow{4}{*}{$10^{6} M_\odot$} \\[10pt] 
\cline{3-5}
 & & $51.8\pm2.3$ & $10.03<b_0\leq 13.16$ & $12.04<b_0\leq 14.22$ &\\[10pt]
 \cline{2-5}
 & \multirow{2}{2.5cm}{$7935$ pc  (Keck team)}& $48.7\pm 7$ & $8.11<b_0\leq 12.99$ & $9.72<b_0\leq 14.03$ &\\[7pt] \cline{3-5}
 & & $51.8\pm 2.3$ & $9.62<b_0\leq 12.61$&$11.54<b_0\leq 13.63$ & \\[10pt]
 \hline
\end{tabular}
\caption{Limit on $b_0$ from the angular diameter of shadows of M87$^*$ and Sgr A$^*$}
\label{table1}
\end{table}

\noindent  It must be mentioned here, that in the absence of a known rotating version of our regular black hole line element, we are forced to make comparisons of EHT results with 
the `theoretically obtained purely circular shadow'. This, surely, may not be fully
accurate since observed deviations from circularity (though small)
(see eg. \cite{cc}) as well
as dependence on the angle of incidence \cite{takahashi} are left out in  our considerations. However, it is also true that the rather ancient (1966) Synge formula \cite{synge} for the shadow  of a simple Schwarzschild
black hole does give fairly good estimates for the average 
shadow radius in M87* and Sgr A*. In the same vein, we may 
be tempted to believe that
our parameter estimates presented here, will not really vary drastically 
if we consider working with a possible rotating generalisation of our spacetime. For a rotating line element, the extra rotation parameter will have to be estimated from observations on `deviations from circularity'.

\section{Conclusion}\label{V}
\noindent We will now review our findings and make a few remarks to wrap up our work. 
\begin{enumerate}
    \item We discuss the GR singularity problem and propose a novel spacetime model which, based on the metric parameters, corresponds to singular, regular black holes or horizon-less compact objects. The line element we work with is restated here:
    \begin{eqnarray}\nonumber\label{general metric} &&ds^{2} = -\Big(1-\frac{b_{0}^{2}r^{2}}{(r^{2}+g^{2})^{2}}\Big)dt^{2} + \Big(1-\frac{b_{0}^{2}r^{2}}{(r^{2}+g^{2})^{2}}\Big)^{-1}dr^{2}  +r^{2}\Big(d\theta^{2}+\sin^{2}\theta d\phi^{2}\Big).
   \end{eqnarray}
   Throughout our discussion, we used a specific parametrization, namely $m^2=g^2/b_0^2$. Depending on the values of $m^2$, different classes of geometries can be constructed from the model mentioned above. For $m^2=0$, we have a mutated Reissner–Nordström geometry with an imaginary charge and a vanishing mass parameter. A family of regular black holes having two horizons can be presented when $0<m^2\leq0.25$.  When $0.25<m^2\leq0.296$ we have spacetimes representing horizon-less compact objects having photon spheres. 
   \item The singular RN-type metric $(m^2=0)$ violates all the classical energy conditions. However, after regularization, the family of regular black holes $(0<m^2\leq0.25)$ seem to satisfy all the energy conditions in a specific range of $r$, though not everywhere. Further, for the regular black hole family, we have shown that components of the curvature tensor and different curvature scalars are indeed finite everywhere.  Following a different path \cite{Modesto}, we have also confirmed that the regular black hole family is geodesically complete for massive and massless test particles. 
   
   \item We found that the family of regular black holes and the horizonless compact objects can be interpreted as the gravitational field sourced by a minimally coupled nonlinear magnetic monopole. The total stress energy can be divided into two parts; one satisfying  NEC and WEC and the other violating them. The source of the singular RN-type metric $(m^2=0)$ may be chosen as a linear magnetic monopole.  As we mentioned in the Introduction, it remains a concern that the source of each regular black hole requires modelling using a different nonlinear electrodynamics Lagrangian. This deficiency needs to be remedied through further explorations in future. We have also shown how the matter required for our geometry may also be modelled using braneworld gravity.
   
   \item Finally, we obtain the circular shadow radius of our regular black hole. Comparing with the observed EHT collaboration results on the average shadow radius, we find that both M 87$^{*}$ and Sgr A$^{*}$ can 
   possibly be modelled by our regular black hole corresponding to particular values for the set $(b_{0}^{2}, m^{2})$. Interestingly, one can also model the data by a horizon-less compact object ($0.25<m^{2}\leq0.296$). It is a fact that a large
   number of astrophysical objects (including hypothetical ones) seem to be possible
   candidates for modeling the shadow images observed till date. One can break this
   `degeneracy' only if we have more observations and we are able to decide which
   of the models have a statistically significant  match with available data. 
   In this paper, we have, however, not tried any such comparisons--our purpose is
   to just demonstrate the feasibility of our model spacetime as a possible candidate.
   
\end{enumerate}
In conclusion, we have found regular spacetimes with and without horizons, sourced by a nonlinear magnetic monopole, which may be used to model the supermassive compact objects as well as ordinary compact objects, known to be abundant in different galaxies in the universe.

\noindent It may be possible to extend our work 
along two directions. One may consider line elements where a Bardeen--like regularising term is added in the metric functions $-g_{tt}$ or $(g_{rr})^{-1}$. The consequences of such an addition may be explored. We can also see if the ultrastatic spacetimes ($g_{tt}=-1$), which arise assuming the same $g_{rr}$ have nonsingular wormhole features. The status of energy conditions and, more importantly, the question of viable matter sources need to be addressed in such scenarios in order to facilitate a proper understanding. Numerous 
suggestions on validating the existence of wormholes through observations have appeared in the literature recently \cite{obworm1,obworm2,obworm3,obworm4} and any study on newer wormhole geometries may therefore have some relevance.
The other direction, as stated before, is about obtaining a rotating version of our spacetime, possibly using the Newman-Janis algorithm \cite{nj}.
Such a rotating spacetime will enable us to make more accurate
estimates of the metric parameters using
observational data.

 \noindent Lastly, an important issue about ours, as well as any spacetime, is its stability under fluctuations. Our preliminary
investigations (not discussed here) indicate that linear perturbations (quasinormal modes) appear to be finite and linear stabilty holds. However, the more
involved question on nonlinear stability needs to be addressed and analysed. 
We hope to return to some of these issues in future work.

\section*{Acknowledgements}
\noindent AK expresses gratitude to Poulami Dutta Roy, Soumya Jana and Pritam Banerjee for their valuable inputs during various discussions. He also thanks Indian Institute of Technology Kharagpur, India, for support through a fellowship and for allowing him to use available computational facilities.

\end{document}